\documentclass[
	twocolumn,
	secnumarabic,
	amsmath,
	amssymb,
	nobibnotes,
	aps,
	pra,
	reprint,
	10pt
]{revtex4-1}

\usepackage{microtype}
\usepackage{mathtools}
\usepackage{array}
\usepackage{booktabs}
\usepackage{dcolumn}
	\newcolumntype{d}[1]{D{.}{.}{#1}}
\usepackage{braket}

\begin{document}

\title{Open systems with error bounds: spin-boson model with spectral density variations}

\author{F. Mascherpa}
\author{A. Smirne}
\author{S. F. Huelga}
\email[]{susana.huelga@uni-ulm.de}
\author{M. B. Plenio}.
\email[]{martin.plenio@uni-ulm.de}
\affiliation{Institut f\"ur Theoretische Physik, Universit\"at Ulm, D-89069, Germany}
\date{\today}%

\begin{abstract}
In the study of open quantum systems, one of the most common ways to describe environmental effects on the reduced dynamics is through the spectral density. However, in many models this object cannot be computed from first principles and needs to be inferred on phenomenological grounds or fitted to experimental data. Consequently, some uncertainty regarding its form and parameters is unavoidable; this in turn calls into question the accuracy of any theoretical predictions based on a given spectral density. Here, we focus on the spin-boson model as a prototypical open quantum system, and find two error bounds on predicted expectation values in terms of the spectral density variation considered, and state a sufficient condition for the strongest one to apply. We further demonstrate an application of our result, by bounding the error brought about by the approximations involved in the Hierarchical Equations of Motion resolution method for spin-boson dynamics.
\end{abstract}

\maketitle

\paragraph*{Introduction ---}

One of the most fundamental models of open quantum systems is the spin-boson model, which comprises a two-level system, such as a spin-$1/2$ particle, and a large number of quantum harmonic oscillators linearly coupled to it and acting as the environment~\cite{CaldeiraLeggettLongArticle, SpinBoson_Leggett, BreuerPetruccione, Weiss}. The influence of these degrees of freedom on the dynamics of the spin can be computed from the strength of the couplings between the spin and each oscillating mode and the frequency of the modes; these quantities can be combined to determine the spectral density of the environment, a function of frequency closely related to its internal correlations and their effect on the reduced dynamics of the spin. Depending on the number of harmonic oscillators present in the model and their dispersion, the spectral density may be a continuous function (for an uncountably infinite set of oscillators) or a linear combination of Dirac delta functions centered at some particular frequencies (for a finite or countably infinite set); the former type is convenient for analytical treatments, while the latter is necessary when performing numerical studies. Generally speaking, the spectral density is not a fundamental object, but rather a phenomenological quantity obtained by making assumptions on the kind of system under study or by fitting experimental data; hence, it is wise to keep in mind that there may always be some error in the functional form considered. This raises the question of how accurate any predictions for a given model can be, given the uncertainty in the spectral density of its environment.

To the best of our knowledge, it appears that no general, rigorous error bound to theoretical predictions for the spin dynamics with respect to changes in the spectral density exists in the literature. The purpose of this work is to address this issue and derive an upper bound to the deviation of the time-dependent expectation value of some spin observable $\hat{O}$ when the spectral density of the oscillator bath changes by a known amount. Aside from the mathematical motivation, such a result would be desirable from a physical point of view for two main reasons: first, when using spectral densities obtained from experiment, it would give a quantitatively certified range for theoretical results to be compatible with them, which can be helpful in order to determine the physical soundness of the theoretical models used; second, it would make it possible to bound the error associated with numerical solutions for the spin-boson model, in analogy with e.~g.\ error bounds on the Time-Evolved Density using Orthogonal Polynomials Algorithm (TEDOPA)~\cite{ChainBounds_WoodsCramerPlenio, ChainBounds_WoodsPlenio}, whenever the method used entails some degree of approximation to the original spectral density of the problem at hand.

We derive our error bounds in the coherent-state path-integral formalism~\cite{NegeleOrland, Kleinert} using the Feynman--Vernon influence functional~\cite{FeynmanVernonFunctional, CaldeiraLeggett}: the idea behind this approach is to treat variations of the spectral density analytically with functional methods, without relying on approximations or numerical techniques. The final results are expressed in terms of canonical quantities such as the interaction Hamiltonian and the bath correlation function, with no need to refer to the path-integral expressions used in the derivation. We will state two forms of the bound, one stronger than the other at long times, and give a sufficient condition for the strong bound to apply, as well as a few examples of spectral density variations complying with it. Finally, we will apply our result to the well-known Hierarchical Equations of Motion (HEOM) resolution method~\cite{HEOM_TanimuraKubo, HEOM_MeierTannor} for spin-boson dynamics.

\paragraph*{The model ---}

Consider the spin-boson Hamiltonian~\cite{SpinBoson_Leggett}
\begin{equation}\label{eq:H}
	\begin{split}
		\hat{H}&=\hat{H}_\mathrm{S}\otimes\mathbb{I}_\mathrm{B}
		+\mathbb{I}_\mathrm{S}\otimes\hat{H}_\mathrm{B}+\hat{H}_\mathrm{I}
		\\
		&=\left(\frac{\epsilon}{2}\sigma_z+\frac{\Delta}{2}\sigma_x\right)\otimes
		\mathbb{I}_\mathrm{B}+\mathbb{I}_\mathrm{S}\otimes\int^\infty_0\!\!\mathrm{d}k\,
		\omega_k\hat{a}^\dagger_k\hat{a}_k
		\\
		&\quad+\frac{\lambda}{2}\sigma_z\otimes\int^\infty_0\!\!\mathrm{d}k\,h(k)
		(\hat{a}^\dagger_k+\hat{a}_k)
	\end{split}
\end{equation}
where $\hat{a}_k$ and $\hat{a}^\dagger_k$ are bosonic creation and annihilation operators satisfying the commutation relation $[\hat{a}_k,\hat{a}^\dagger_l]=\delta_{kl}$, the continuous label $k$ identifying each boson may be thought of as a momentum variable and $\omega_k$ as a dispersion relation, which we take to be linear: $\omega_k=gk$, with unit $g$ and in natural units with $\hbar=1$; the function $h(k)$ expresses the coupling strength between each mode and the qubit. Depending on the choice of $h$ (which may well include delta functions), the model may have a finite, countably infinite or uncountably infinite number of bosons, with $k$ bounded or unbounded; we will always assume the domain of $h$ to be the whole positive $k$ axis. With $\omega_k=k$, there is a one-to-one correspondence between the choice of $h(k)$ and the spectral density $J(\omega)$ of the bosonic environment: $J(\omega_k)=\pi h^2(\omega_k)$~\cite{ChainMapping_PriorPlenio, ChainMapping_WoodsPlenio, ChainMapping_ChinPlenio}.

To complete our ansatz, we take the initial state to be of the form $\hat{\rho}_0=\hat{\rho}_{\mathrm{S}0}\otimes\frac{e^{-\beta\hat{H}_\mathrm{B}}}{\mathrm{Tr}_\mathrm{B}\left(e^{-\beta\hat{H}_\mathrm{B}}\right)}$, where $\hat{\rho}_{\mathrm{S}0}$ is arbitrary and the bosons are in thermal equilibrium at temperature $T=\frac{1}{k_\mathrm{B}\beta}$. In principle, this assumption could be relaxed to include more general Gaussian initial states for the bath~\cite{SpinBoson_Leggett}, such as a thermal state perturbed by a laser pulse before the interaction with the spin begins, but for the sake of simplicity we will keep our treatment within the standard framework of thermal environments for the time being, and leave extensions to this first model to our future work.

Under these assumptions, the expectation value $\langle \hat{O}(t)\rangle$ of some spin observable $\hat{O}$ at time $t$
\begin{equation}\label{eq:ExpVal}
	\langle \hat{O}(t)\rangle=\mathrm{Tr}\left(\hat{O}e^{-i\hat{H}t}
	\hat{\rho}_0e^{i\hat{H}t}\right)
\end{equation}
depends on the environment only via $J(\omega)$~\cite{CaldeiraLeggett}. We will use functional analysis tools in order to quantify and bound the dependence of $\langle \hat{O}(t)\rangle$ on $J$: in order to better emphasize this concept, in what follows we shall refer to expectation values specifying the relevant spectral density as a subscript, effectively regarding $\langle \hat{O}(t)\rangle_J$ as a functional on the space of spectral densities as well as a function of time.

\paragraph*{Spectral density variations and error bounds ---}

We want to compare the expectation values of $\hat{O}$ for arbitrary spectral densities $J_0(\omega)$ and $J(\omega)\coloneqq J_0(\omega)+\Delta J(\omega)$: in other words, we are interested in bounding the absolute value of the difference
\begin{equation}\label{eq:DeltaExpVal}
	\Delta\langle\hat{O}(t)\rangle\coloneqq\langle\hat{O}(t)\rangle_J
	-\langle\hat{O}(t)\rangle_{J_0}.
\end{equation}
The path integral formalism~\cite{CaldeiraLeggett, SpinPathIntegral_Radcliffe, SpinPathIntegral_Klauder, SpinPathIntegral_KuratsujiSuzuki, SpinPathIntegral_Kochetov, SpinPathIntegral_Kirchner, CoherentPathIntegral_KordasMistakidisKaranikas, SpinPathIntegral_KordasKalantzisKaranikas} makes it easy to eliminate the bosonic degrees of freedom from the expression for the expectation value of $\hat{O}(t)$ by performing the relevant Gaussian integral analytically (See Appendix~\ref{sec:CSPI}). The result is a path integral for the spin variables alone, with the time evolutions of the left and right part of the initial state no longer independent. The Feynman--Vernon influence functional encodes this mixing, which is a result of the partial trace over the bath: it has the form of a Gaussian functional of the spin variables, with the bath correlation function
\[
	\xi_J(t)\coloneqq\int^\infty_0\!\!\frac{\mathrm{d}\omega}{\pi}\,
	J(\omega)\left(\coth\left(\frac{\beta\omega}{2}\right)\cos(\omega t)+
	i\sin(\omega t)\right)
\]
coupling them. Note that since we have assumed a thermal initial state for the bosons, which is stationary with respect to their free dynamics, $\xi_J(t)$ is not a function of two time variables, but merely of their difference.

To write out $\langle \hat{O}(t)\rangle_J$ explicitly in terms of $J_0(\omega)$ and $\Delta J(\omega)$, define the Heisenberg-picture operator $\hat{h}_\mathrm{I}(t)\coloneqq e^{i\hat{H}t}\left(\frac{\lambda}{2}\sigma_z\otimes\mathbb{I}_\mathrm{B}\right)e^{-i\hat{H}t}$
and the super-operator
\begin{multline}\label{eq:InfluenceSuperOperator}
	\hat{\Phi}[\hat{h}_\mathrm{I},\hat{h}'_\mathrm{I},J]\coloneqq\mathcal{T}
	\int_0^t\!\!\mathrm{d}t'\int_0^{t'}\!\!\mathrm{d}t''\,
	\left(\hat{h}_\mathrm{I}(t')-\hat{h}'_\mathrm{I}(t')\right)
	\\
	\left(\xi_J(t'-t'')\hat{h}_\mathrm{I}(t'')-\xi^*_J(t'-t'')\hat{h}'_\mathrm{I}(t'')
	\right),
\end{multline}
which acts on a spin state $\hat{\rho}_0$ with all $\hat{h}_\mathrm{I}(t)$ operators multiplying it from the left and all $\hat{h}'_\mathrm{I}(t)$ from the right in the appropriate time order. This is, up to an overall minus sign, the operator version of the logarithm of the influence functional, as shown in Appendix~\ref{sec:CSPI}. Using the exponential form of the Feynman--Vernon functional and the fact that $\hat{\Phi}[\hat{h}_\mathrm{I},\hat{h}'_\mathrm{I},J]$ is linear in the spectral density, it can be shown that
\begin{equation}\label{eq:DeltaExpValSeries}
	\Delta\langle\hat{O}(t)\rangle=\sum^\infty_{n=1}\frac{\langle\hat{O}(t)\mathcal{T}
	(-\hat{\Phi}[\hat{h}_\mathrm{I},\hat{h}'_\mathrm{I},\Delta J])^n\rangle_{J_0}}{n!}.
\end{equation}
Note that extending the same series by including the term with $n=0$ just adds $\langle \hat{O}(t)\rangle_{J_0}$, giving $\langle \hat{O}(t)\rangle_J$.

The series in Eq.~\eqref{eq:DeltaExpValSeries} may be bounded in magnitude term by term, using the singular-value decomposition of the spin operators to remove the complicated time dependence of the time-ordered correlation functions, and then summed: the result is the general formula
\begin{equation}\label{eq:GeneralBound}
	|\Delta\langle\hat{O}(t)\rangle|\leq||\hat{O}||
	\left(e^{\lambda^2\int^t_0\!\!\mathrm{d}t'\int^{t'}_0\!\!\mathrm{d}t''\,
	|\Delta\xi(t'-t'')|}-1\right),
\end{equation}
where $\Delta\xi(t)\coloneqq\xi_J(t)-\xi_{J_0}(t)=\xi_{\Delta J}(t)$ and we have used the operator norm $||\hat{O}||\coloneqq||\hat{O}||_\infty=\sigma_1(\hat{O})$, $\sigma_1(\hat{O})$ being the highest singular value of $\hat{O}$.

Depending on $\Delta J(\omega)$, there are two options for bounding the double time integral in Eq.~\eqref{eq:GeneralBound}: the worst-case scenario is a $\Delta\xi(t)$ which never decays, as would be the case for singular contributions such as $\Delta J(\omega)=\kappa\delta(\omega-\omega_0)$. Then one would be forced to bound $|\Delta\xi(t)|$ by some constant $C>0$, obtaining the error bound
\begin{equation}\label{eq:WeakBound}
	\begin{split}
		|\Delta\langle\hat{O}(t)\rangle|\leq
		||\hat{O}||\left(e^{\frac{\lambda^2Ct^2}{2}}-1\right).
	\end{split}
\end{equation}

However, if $\Delta J(\omega)$ is such that the resulting $\Delta\xi(t)$ decays fast enough to be absolutely integrable, i.~e.
\begin{equation}\label{eq:StrongBoundCondition}
	\int^{\infty}_0\!\!\mathrm{d}t\,|\Delta\xi(t)|=c<\infty,
\end{equation}
then one can tighten the bound considerably. This is the case for many physically relevant situations, e.~g.\ for ohmic, superohmic or antisymmetrized Lorentzian spectral density variations; a few relevant examples are given in Appendix~\ref{sec:Examples}. In practice, it is often easier to apply the triangle inequality to $|\Delta\xi(t)|$ first and then bound its real and imaginary parts separately, even though this may weaken the bound slightly: the result is
\begin{equation}\label{eq:StrongBound}
	|\Delta\langle\hat{O}(t)\rangle|\leq||\hat{O}||
	\left(e^{\lambda^2(\gamma(\beta)+\eta)t}-1\right),
\end{equation}
where
\begin{align*}
	\gamma(\beta)&\coloneqq\int^\infty_0\!\!\mathrm{d}t\,\left|\int^\infty_0\!\!
	\frac{\mathrm{d}\omega}{\pi}\,\Delta J(\omega)
	\coth\left(\frac{\beta\omega}{2}\right)\cos(\omega t)\right|
	\\
	\eta&\coloneqq\int^\infty_0\!\!\mathrm{d}t\,\left|\int^\infty_0\!\!
	\frac{\mathrm{d}\omega}{\pi}\,\Delta J(\omega)\sin(\omega t)\right|;
\end{align*}
this is the central result of this Letter.

The error bound~\eqref{eq:StrongBound} manifestly satisfies all properties we expect from it: it is proportional to the norm of the spin observable being evaluated, vanishes at $t=0$ and grows exponentially in time, which makes it scale linearly at short times, at a rate proportional to the square of the coupling in accordance with the relation $J(\omega)=\pi h^2(\omega)$. Note that the norm of the observable itself only enters the result as a prefactor: this is expected because the error is a consequence of an incomplete knowledge of the dynamics of the system, regardless of what observable is being estimated; the relative error bound is thus the same for all observables and only needs to be computed once.

Both bounds are very weak at long times because by construction they keep no account of the free dynamics of the spin. It is worth mentioning, however, that the singular-value decomposition used in our derivation does not affect the bounds in the case of pure dephasing, in which $[\hat{h}_\mathrm{I},\hat{H}_\mathrm{S}]=0$ and no interference effects due to time evolution take place inside $\hat{\Phi}[\hat{h}_\mathrm{I},\hat{h}'_\mathrm{I},J]$: pure dephasing is the worst-case scenario with respect to this inequality.

\paragraph*{Application to Hierarchical Equations of Motion ---}

The HEOM method for solving open-system problems beyond standard perturbation theory was first proposed and tested around 1990 by Kubo, Tanimura and others~\cite{HEOM_TanimuraKubo, HEOM_Tanimura, HEOM_TanimuraWolynes} for antisymmetrized Lorentzian spectral densities $J_L(\omega;\Omega,\Gamma)=\frac{\pi}{2}\frac{\omega}{((\omega+\Omega)^2+\Gamma^2)((\omega-\Omega)^2+\Gamma^2)}$; their scheme replaces the possibly non-Markovian generalized quantum master equation for the state of some open system with a system of time-local differential equations for both the reduced density matrix and a set of so-called auxiliary density matrices, which encode information about the bath. In principle, this hierarchy of equations is infinite, but in computations it is necessary to truncate it at some order, which may be much higher than conventional perturbative approaches can usually attain~\cite{HEOM_LiuZhuBaiShi}.

The form of the bath spectral density is an important part of the scheme, because it is necessary for the bath correlation function to have the form of a sum of exponentials, as is the case with antisymmetrized Lorentzians; however, Meier and Tannor have shown~\cite{HEOM_MeierTannor} that many other spectral densities may be approximated very accurately by a suitable linear combination of Lorentzians, greatly extending the applicability of the method. Later studies such as Ref.~\cite{HEOM_LiuZhuBaiShi} also explored the possibility of fitting arbitrary bath spectral densities using other functions yielding exponentially damped correlations.

We will now apply our findings to the results presented in Ref.~\cite{HEOM_MeierTannor} on the spin-boson application of HEOM; for the details of how the problem is formulated, the interested reader is referred to the original paper. For our purposes, it is sufficient to say that for an antisymmetrized Lorentzian spectral density, which yields a correlation function
\begin{equation}\label{eq:LorentzianCorrelation}
	\begin{split}
		\xi_L(t;\Omega,\Gamma)=&\frac{e^{-\Gamma t}}{8\Omega\Gamma}\left(\coth\left(
		\frac{\beta}{2}(\Omega+i\Gamma)\right)e^{i\Omega t}\right.
		\\
		&\left.+\coth\left(\frac{\beta}{2}(\Omega-i\Gamma)\right)e^{-i\Omega t}
		+2i\sin(\Omega t)\right)
		\\
		&-\frac{2}{\beta}\sum^\infty_{k=1}
		\frac{\nu_ke^{-\nu_kt}}{(\Omega^2+\Gamma^2-\nu^2_k)^2+4\Omega^2\nu^2_k},
	\end{split}
\end{equation}
where $\nu_k\coloneqq\frac{2\pi k}{\beta}$ are the Matsubara frequencies, the scheme computes dynamics and operator expectation values corresponding to a truncation of the series in Eq.~\eqref{eq:LorentzianCorrelation} at order $N$. The accuracy of this approximation is unknown, and one usually performs numerical simulations with increasing $N$ until the results stop changing appreciably. Convergence is thus declared heuristically, assuming that if the distance between the results obtained and those given by $N-1$ is negligible, then so is the difference between them and the true physics given by $\xi(t)$. With our result Eq.~\eqref{eq:StrongBound}, the maximum distance between the predictions for some value of $N$ and the physically correct result at $N\longrightarrow\infty$ may be determined with a few lines of simple algebra instead of running an unpredictable number of costly simulations: we will now demonstrate this by giving the results of our bound~\eqref{eq:StrongBound} applied to the simulations in Meier and Tannor's paper~\cite{HEOM_MeierTannor}.

In their model, the spin Hamiltonian is $\hat{H}_\mathrm{S}\coloneqq\frac{\epsilon}{2}(\sigma_z+\sigma_x)$, the coupling is given by $	\xi\coloneqq\frac{\lambda^2}{4}=0.1$ and the spectral density considered is ohmic and defined as $J(\omega)\coloneqq\frac{\pi}{2}\omega e^{-\omega/\Omega}$ with $\Omega=\frac{15}{4}\epsilon$ and fitted with three Lorentzians whose parameters are listed in Table~\ref{tb:JFitParameters}.

\begin{table}
	\centering
	\begin{tabular*}{.3\textwidth}{@{\extracolsep{\fill}}d{0}d{0}d{0}}
	\toprule
	\multicolumn{1}{c}{$\displaystyle{\frac{p_i}{\xi\Omega^4}}$}
	& \multicolumn{1}{c}{$\displaystyle{\frac{\Omega_i}{\Omega}}$}
	& \multicolumn{1}{c}{$\displaystyle{\frac{\Gamma_i}{\Omega}}$} \\
	\midrule
	$12.0677$ & $0.2378$ & $2.2593$ \\
	$-19.9762$ & $0.0888$ & $5.4377$ \\
	$0.1834$ & $0.0482$ & $0.8099$ \\
	\bottomrule
	\end{tabular*}
	\caption{\label{tb:JFitParameters}Parameters of the reconstructed spectral density $J(\omega)=\sum^3_{i=1}p_iJ_L(\omega;\Omega_i,\Gamma_i)$ from Ref.~\cite{HEOM_MeierTannor}.}
\end{table}

For a general linear combination of antisymmetrized Lorentzians $J(\omega)=\sum^n_{i=1}p_iJ_L(\omega;\Omega_i,\Gamma_i)$, absorbing the overall coupling strength $\lambda^2$ in the coefficients $p_i$ for the sake of simplicity, the truncation of $\xi(t)$ at order $N$ gives
\[
	\Delta\xi(t)=-\frac{\pi}{\beta}\sum^n_{i=1}\sum^\infty_{k=N+1}
	\frac{p_i\nu_ke^{-\nu_kt}}{(\Omega_i^2+\Gamma_i^2-\nu^2_k)^2+4\Omega_i^2\nu^2_k},
\]
which is real and satisfies condition~\eqref{eq:StrongBoundCondition} (see Appendix~\ref{sec:Examples}), and hence
\begin{equation}\label{eq:GammaTriangle}
	\begin{split}
		\gamma(\beta)&=\int^\infty_0\!\!\mathrm{d}t|\Delta\xi(t)|
		\\
		&\leq\frac{\pi}{\beta}\sum^n_{i=1}\sum^\infty_{k=N+1}
		\frac{|p_i|}{(\Omega_i^2+\Gamma_i^2-\nu^2_k)^2+4\Omega_i^2\nu^2_k}
	\end{split}
\end{equation}
and $\eta=0$. The series $\sum^\infty_{k=1}\frac{1}{(\Omega^2+\Gamma^2-\nu^2_k)^2+4\Omega^2\nu^2_k}$ can be summed exactly, so we obtain the result as a difference:
\begin{multline}\label{eq:GammaN}
	\gamma_N(\beta)\coloneqq\frac{\pi}{2\beta}\sum^n_{i=1}|p_i|
	\Bigg(-\frac{1}{(\Omega^2_i+\Gamma^2_i)^2}
	\\
	+\frac{\beta\Omega_i\sin(\beta\Gamma_i)
	+\beta\Gamma_i\sinh(\beta\Omega_i)}{4\Omega_i\Gamma_i
	(\Omega^2_i+\Gamma^2_i)(\cosh(\beta\Omega_i)-\cos(\beta\Gamma_i))}
	\\
	-\sum^N_{k=1}
	\frac{2}{(\Omega^2_i+\Gamma^2_i-\nu^2_k)^2+4\Omega^2_i\nu^2_k}\Bigg),
\end{multline}
using the triangle inequality on the $|p_i|$ as in Eq.~\eqref{eq:GammaTriangle}.

In Ref.~\cite{HEOM_MeierTannor}, the authors computed the time evolution of the expectation value $\langle\sigma_z\rangle$ at temperatures $\epsilon\beta=0.4$, $1.4$ and $10.0$ for times until $\epsilon t_\mathrm{max}=30$, at which point the system has thermalized almost completely. The number $N$ of Matsubara frequencies needed for convergence for these three temperatures was $2$, $7$ and $48$ respectively, due to the better performance of the HEOM method at high temperatures.

We calculated the error bound for all three cases, both with Eq.~\eqref{eq:GammaN} and by performing the integral in Eq.~\eqref{eq:GammaTriangle} numerically instead of using the triangle inequality; in order to better assess the quality of our bound, we have also determined the truncation order necessary for the maximum error given by either bound to drop below $20\%$ at each temperature. Table~\ref{tb:Error} shows our results.

\begin{table}
	\centering
	\begin{tabular*}{.48\textwidth}{@{\extracolsep{\fill}}d{0}cd{5}d{5}cc}
	\toprule
	\multicolumn{1}{c}{$\displaystyle{\epsilon\beta}$}
	& \multicolumn{1}{c}{$\displaystyle{N}$}
	& \multicolumn{1}{c}{$\displaystyle{\frac{|\Delta\langle\sigma_z\rangle(t_\mathrm{max})|}{||\sigma_z||}^\mathrm{an}\!\!\!\!(N)}$}
	& \multicolumn{1}{c}{$\displaystyle{\frac{|\Delta\langle\sigma_z\rangle(t_\mathrm{max})|}{||\sigma_z||}^\mathrm{num}\!\!\!\!\!\!\!(N)}$}
	& \multicolumn{1}{c}{$\displaystyle{N^\mathrm{an}_{20\%}}$}
	& \multicolumn{1}{c}{$\displaystyle{N^\mathrm{num}_{20\%}}$} \\
	\midrule
	$0.4$ & $2$ & $27.94\%$ & $9.43\%$ & $3$ & $2$ \\
	$1.4$ & $7$ & $62.39\%$ & $23.77\%$ & $10$ & $8$ \\
	$10.0$ & $48$ & $111.69\%$ & $45.34\%$ & $70$ & $56$ \\
	\bottomrule
	\end{tabular*}
	\caption{\label{tb:Error}Results for the analytical and numerical bounds on the relative error at time $t_\mathrm{max}$, for the three cases considered in the original paper Ref.~\cite{HEOM_MeierTannor}. The last two columns indicate at what $N$ the maximum relative error from both calculations would be under $20\%$.}
\end{table}

The numerical integral gives remarkably strong bounds at the timescale of interest, given the exponential time dependence of our result Eq.~\eqref{eq:StrongBound}: the maximum difference between the predicted and the actual value of $\langle\sigma_z\rangle$ at time $t_\mathrm{max}$ is guaranteed to lie between $0.09||\sigma_z||=0.09$ and $0.46||\sigma_z||=0.46$ in all three cases and $\gamma(\beta)$ is small enough for the time scaling to be well within the linear regime at time $t_\mathrm{max}$, which is of the order of the equilibration time of the system~\cite{HEOM_MeierTannor}. It should also be noted that in many relevant applications (e.~g.\ transient spectroscopy) the timescales of interest are much shorter.

Because the coefficients $p_i$ of the components of the fitted spectral density and correlation function are both positive and negative while the analytical formula~\eqref{eq:GammaN} only uses their absolute values, it overestimates $|\Delta\xi(t)|$ and $\gamma(\beta)$ considerably, explaining the suboptimal results given by the fully analytical bound for the case at hand.

\paragraph*{Conclusions ---}

We have investigated the sensitivity of spin operator expectation values in the spin-boson model to changes in the spectral density, and derived two rigorous time-dependent error bounds under the only assumptions of factorizing initial conditions and a linearly coupled thermal bath of quantum harmonic oscillators. The results depend on the system-bath coupling strength and the spectral density variation considered, and can be expressed in a simple and elegant form in terms of these quantities. We also found the encouraging result that most of the commonly used bath models obey the strongest of the two bounds, the exceptions being baths with slowly- or non-decaying correlation functions.

These error bounds may be applied in many physically relevant contexts, such as comparing theoretical predictions with experimental results based on spectral densities known up to some error, determining whether a given environmental spectrum constitutes a reasonable ansatz for a physical system for which experimental or numerical data are available, or certifying the accuracy of theoretical or numerical results obtained by changing the bath correlation function in order to solve for the dynamics.

As an example application, we have demonstrated the latter use of the error bound by applying it to existing numerical results obtained with the HEOM scheme: we have shown that our results can quantitatively certify the robustness of the method, providing useful bounds on the maximum physically possible difference between the predicted and the exact results, and that it can therefore be used to ascertain the achieved precision without testing it against more costly numerical computations.

In addition to backing up theoretical predictions with rigorous error bounds and finding practical applications in computational contexts such as HEOM simulations, this work also provides a route for the derivation of analogous bounds on many-time correlation functions or open quantum systems more complex than the spin-boson model, such as $n$-level systems, spin chains or the like, as long as the environment and initial conditions satisfy the same assumptions and bounded observables are considered.

The authors thank James Lim for useful discussions about HEOM. This work was supported by an Alexander von Humboldt Professorship, the ERC Synergy grant BioQ, the CRC/TR21, the H2020- FETPROACT-2014 Grant QUCHIP (Quantum Simulation on a Photonic Chip; GA  641039, http://www.quchip.eu) and the FP7 project PAPETS, GA 323901. 

\appendix

\section{\label{sec:CSPI}Coherent-state path integral for the spin-boson model and derivation of the error bounds}

In this appendix, we will rephrase the spin-boson problem in terms of both bosonic and spin coherent-state path integrals, remove the bosons from the problem and give the form of the influence functional for the qubit in terms of the spectral density, and use this result to derive Eq.~\eqref{eq:GeneralBound} of the main text.

\subsection{\label{sec:CSPI-Bosons}Bosonic coherent-state path integral}

For a single bosonic degree of freedom with creation and annihilation operators $\hat{a}^\dagger$ and $\hat{a}$, define an unnormalized coherent state by
\begin{equation}
	\ket{\phi}\coloneqq e^{\phi\hat{a}^\dagger}\ket{0},
\end{equation}
where $\ket{0}$ is the vacuum state and $\phi$ is any complex number. This state is an eigenstate of $\hat{a}$ with eigenvalue $\phi$; its Hermitian conjugate $\bra{\phi}=\bra{0}e^{\phi^*\hat{a}}$ is a left eigenstate of $\hat{a}^\dagger$ with eigenvalue $\phi^*$. The overlap between any two such states $\ket{\phi}$ and $\ket{\phi'}$ is
\[
	\braket{\phi|\phi'}=\bra{0}e^{\phi^*\hat{a}}e^{\phi'\hat{a}^\dagger}\ket{0}
	=e^{\phi^*\phi'}.
\]

Coherent states form an overcomplete Hilbert space basis; the relevant closure relation is
\begin{equation}
	\int\frac{\mathrm{d}\phi^*\mathrm{d}\phi}{2\pi i}\,e^{-|\phi|^2}\ket{\phi}\bra{\phi}
	=\mathbb{I},
\end{equation}
where the factor $e^{-|\phi|^2}$ compensates for the fact that the squared norm of $\ket{\phi}$ is $\braket{\phi|\phi}=e^{|\phi|^2}$. Our notation for the integral measure is related to that used in other papers by
\[
	\frac{\mathrm{d}\phi^*\mathrm{d}\phi}{2\pi i}
	=\frac{\mathrm{d}\Re(\phi)\mathrm{d}\Im(\phi)}{\pi},
\]
where $\Re(\phi)$ and $\Im(\phi)$ denote the real and imaginary parts of $\phi$.

The generalization to multiple degrees of freedom is straightforward and gives coherent states $\ket{\Phi}$ specified by their eigenvalues $\phi_k$ for each annihilation operator $\hat{a}_k$; the closure relation in this case reads 
\begin{equation}\label{eq:ClosureBosons}
	\int\prod_k\frac{\mathrm{d}\phi^*_k\mathrm{d}\phi_k}{2\pi i}\,
	e^{-\sum_k|\phi_k|^2}\ket{\Phi}\bra{\Phi}=\mathbb{I}.
\end{equation}

Consider the propagator from some initial coherent state $\ket{\Phi_i}$ at time~$0$ to a final coherent state $\ket{\Phi_f}$ at time~$t$: $\bra{\Phi_f}e^{-i\hat{H}t}\ket{\Phi_i}$ (with $\hat{H}=\sum_k\omega_k\hat{a}^\dagger_k\hat{a}_k$). Performing the Trotter decomposition of the time evolution, inserting the resoution of the identity~\eqref{eq:ClosureBosons} at every intermediate step and taking the continuum limit, one obtains the path integral
\begin{multline}\label{eq:BosonPropagator}
	\bra{\Phi_f}e^{-i\hat{H}t}\ket{\Phi_i}=
	\int^{\phi_k^*(t)=\phi^*_{kf}}_{\phi_k(0)=\phi_{ki}}\!\!
	\mathcal{D}\mu(\phi_k,\phi^*_k)
	\\
	e^{\Gamma[\phi_k,\phi^*_k]+i\mathcal{S}[\phi_k,\phi^*_k]},
\end{multline}
where the variables $\phi_k$, $\phi^*_k$ have become time-dependent fields and we have absorbed the denominators $2\pi i$ into the definition of the symbolic functional measure $\mathcal{D}\mu(\phi_k,\phi^*_k)$. The first term in the exponent
\[
	\Gamma[\phi_k,\phi^*_k]\coloneqq\frac{1}{2}\sum_k
	\left(\phi^*_{kf}\phi_k(t)+\phi_k^*(0)\phi_{ki}\right)
\]
is a boundary term (note that the boundary conditions in~\eqref{eq:BosonPropagator} fix $\phi_k(0)$ and $\phi^*_k(t)$, but not $\phi^*_k(0)$ and $\phi_k(t)$, which are separate, independent variables~\cite{NegeleOrland, Kleinert}), and the action $\mathcal{S}[\phi_k,\phi^*_k]$ is
\begin{multline*}
	\mathcal{S}[\phi_k,\phi^*_k]\coloneqq\int^t_0\!\!\mathrm{d}t'\,
	\left(\frac{i}{2}\sum_k\left(\phi^*_k(t')\dot{\phi}_k(t')\right.\right.
	\\
	\left.-\dot{\phi}^*_k(t')\phi_k(t')\right)
	-H\left(\phi^*_k(t'),\phi_k(t')\right)\Bigg)
\end{multline*}
where $H\left(\phi^*_k(t'),\phi_k(t')\right)\coloneqq\sum_k\omega_k\phi^*_k(t')\phi_k(t')$.

The propagator~\eqref{eq:BosonPropagator} is the result of averaging two different prescriptions for the continuum limit, which would otherwise give just one time-derivative term in the action and either the initial or the final part of the boundary term $\Gamma[\phi_k,\phi^*_k]$ (see e.~g.\ Negele and Orland~\cite{NegeleOrland}). Both choices yield correct results if viewed as mere formal expressions for the underlying discrete path integral; however, the symmetrized prescription we have used has the advantage of giving an object which can be consistently used in the continuum without referring to the discrete expression, making it easier and more natural to work with~\cite{Kleinert, CoherentPathIntegral_KordasMistakidisKaranikas}.

\subsection{\label{sec:CSPI-Spin}Spin coherent-state path integral}

The state space of a qubit is spanned by the `up' and `down' orthogonal spin states $\ket{\uparrow}$ and $\ket{\downarrow}$, which are mapped onto each other by the raising and lowering operators $\sigma^+$ and $\sigma^-$:
\begin{align}
	\sigma^+\ket{\downarrow}&=\ket{\uparrow},\quad\sigma^-\ket{\uparrow}=\ket{\downarrow}
	\\
	\sigma^+\ket{\uparrow}&=\sigma^-\ket{\downarrow}=0.
\end{align}

There are two equivalent definitions of spin coherent states, based on the choice of $\ket{\uparrow}$ or $\ket{\downarrow}$ as a reference state analogous to $\ket{0}$ in the bosonic case. Taking $\ket{0}\coloneqq\ket{\downarrow}$ (and using the notation $\ket{1}$ for $\ket{\uparrow}$ from now on), define the normalized coherent state
\begin{equation}
	\ket{z}\coloneqq\frac{1}{\sqrt{1+|z|^2}}e^{z\sigma^+}\ket{0}
\end{equation}
where $z\in\mathbb{C}$. For general representations of $SU(2)/U(1)$ with spin $s\neq\frac{1}{2}$, an analogous definition applies, with the square root at the denominator replaced by the power of $s$ in order to keep the states normalized~\cite{SpinPathIntegral_Radcliffe, SpinPathIntegral_Klauder, SpinPathIntegral_KuratsujiSuzuki, SpinPathIntegral_Kochetov}.

The states are clearly overcomplete, since no coherent state defined this way can be orthogonal to $\ket{0}$, and the overlap between two coherent states $\ket{z}$ and $\ket{z'}$ is
\[
	\braket{z|z'}=\frac{1+z^*z'}{\sqrt{(1+|z|^2)(1+|z'|^2)}};
\]
the resolution of the identity for spin coherent states is
\begin{equation}\label{eq:ClosureSpin}
	\int\frac{\mathrm{d}z^*\mathrm{d}z}{2\pi i}\,\frac{2}{(1+|z|^2)^2}\ket{z}\bra{z}
	=\mathbb{I}.
\end{equation}

Geometrically, this definition is related to the stereographic projection of the Bloch sphere: the complex parameter $z=\tan\left(\frac{\theta}{2}\right)e^{i\phi}$, where $\theta\in[0,\pi]$ and $\phi\in[0,2\pi)$, uniquely determines the Bloch vector corresponding to the state $\ket{z}$ through
\begin{align}
	\mathbf{n}(z)&=\left(\frac{2\Re(z)}{1+|z|^2},\frac{2\Im(z)}{1+|z|^2},
	\frac{|z|^2-1}{1+|z|^2}\right),
	\\
	z(\mathbf{n})&=\frac{n_x+in_y}{1-n_z}
\end{align}
and the measure in~\eqref{eq:ClosureSpin} gives the area element on the Bloch sphere. Note that in the limit $|z|\longrightarrow\infty$ the Bloch vector approaches $\mathbf{n}(\infty)=(0,0,1)$, which corresponds to $\ket{1}$, regardless of the phase of $z$, emphasizing the isomorphism between $SU(2)/U(1)$, the unit sphere and the one-point compactified complex plane.

The propagator from a state $\ket{z_i}$ at time~$0$ to a state $\ket{z_f}$ at time~$t$ with the dynamics given by a Hamiltonian $\hat{H}$ is
\begin{equation}
	\bra{z_f}e^{-i\hat{H}t}\ket{z_i}
	=\int^{z^*(t)=z^*_f}_{z(0)=z_i}\!\!\mathcal{D}\mu(z,z^*)\,
	e^{\Gamma[z,z^*]+i\mathcal{S}[z,z^*]},
\end{equation}
where $\mathcal{D}\mu(z,z^*)$ is the path-integral measure, which includes all factors in~\eqref{eq:ClosureSpin}, $\Gamma[z,z^*]$ is a boundary term and $\mathcal{S}[z,z^*]$ the effective action for the fields $z$ and $z^*$. The structure of these terms is
\[
	\Gamma[z,z^*]\coloneqq\frac{1}{2}\log\left(
	\frac{(1+z^*(0)z_i)(1+z^*_fz(t))}{(1+|z_i|^2)(1+|z_f|^2)}\right),
\]
where again only $z(0)$ and $z^*(t)$ are fixed by the boundary conditions, but not $z^*(0)$ and $z(t)$,
and
\begin{multline*}
	\mathcal{S}[z,z^*]\coloneqq\int_0^t\!\!\mathrm{d}t'\,
	\left(\frac{i}{2}\frac{z^*(t')\dot{z}(t')-\dot{z}^*(t')z(t')}{1+|z(t')|^2}\right.
	\\
	-H\left(z(t'),z^*(t')\right)\bigg).
\end{multline*}

The term with the time derivatives of the fields has a geometrical interpretation as an external differential on the spherical state manifold, and gives rise to a Berry phase in closed-contour path integrals such as the partition function for the spin: this Berry-phase term was found to have relevant implications in certain problems, such as the quantum-to-classical mapping of the phase transition theory for the subohmic spin-boson model~\cite{SpinPathIntegral_Kirchner}.

In general, determining the form of $H\left(z(t'),z^*(t')\right)$ is a nontrivial task. Many papers define it as $\bra{z(t')}\hat{H}\ket{z(t')}$, the limit of the object $\bra{z_{i+1}}\hat{H}\ket{z_i}$ up to leading order in the timestep $\Delta t$ as $\Delta t\longrightarrow 0$ (see e.~g.\ the derivations in Refs.~\cite{SpinPathIntegral_KuratsujiSuzuki} or~\cite{SpinPathIntegral_Kochetov}). However, calculations using this form of $H\left(z(t'),z^*(t')\right)$ in the continuum limit have long been known to give inconsistent results for some systems~\cite{SpinPathIntegral_Kochetov, SpinPathIntegral_KordasKalantzisKaranikas}. It has often been said that the blame for this lies in the continuum limit being merely formal and not to be taken as a mathematically legitimate operation, and that the correct way to deal with doubtful situations is to perform all calculations in the discrete case and take the continuum limit only at the end. This is tantamount to deriving each result from scratch instead of using the continuous path integral as a convenient, reliable and well-established tool, which defeats the purpose of using path integrals in the first place, and is arguably one of the reasons for the failure of $SU(2)$ coherent-state path integrals to attain quite the same popularity and widespread use as more standard techniques routinely and safely employed in quantum and statistical mechanics or quantum field theory.

However, Kordas and coworkers have shown in recent work~\cite{SpinPathIntegral_KordasKalantzisKaranikas} that this is not the case, and described a prescription for deriving $H\left(z(t'),z^*(t')\right)$ from $\hat{H}$ which consistently yields correct results. They point out that the object weighting the paths in the standard Feynman path integral is the classical action of the system, and that an analogous object should be sought for spin problems as well, a subtlety which is not so apparent for infinite-dimensional systems. Applying the Holstein--Primakoff transformation to the $SU(2)$ representation at hand in order to define effective creation and annihilation operators, they proceed to express $\hat{H}$ in terms of the effective position and momentum operators associated with them, and then take the expectation value of these degrees of freedom on the state $\ket{z(t')}$; the result coincides with $\bra{z(t')}\hat{H}\ket{z(t')}$ in the $s\longrightarrow\infty$ limit and for $s=\frac{1}{2}$, which is the case we are interested in, but generally differs from it by $s$-dependent terms.

\subsection{\label{sec:CSPI-InfFunc}Influence functional for the spin-boson problem}

The main reason for introducing coherent-state path integrals for our system is that it is necessary to use path integrals over continuous bases in order to take the continuum limit in an unambiguous way~\cite{SpinPathIntegral_Kirchner}, and hence consistently perform the Gaussian integration of the bosonic fields. With the definitions given above, we now rephrase the original problem in the path-integral language and eliminate the bosons in favor of the influence functional, from which we will derive the results presented in the main text.

The object $\langle \hat{O}(t)\rangle_J=\mathrm{Tr}\left(\hat{O}e^{-i\hat{H}t}\hat{\rho}_0e^{i\hat{H}t}\right)$ takes the following path-integral form:
\begin{widetext}
	\begin{multline}\label{eq:ExpValFullPI}
		\langle \hat{O}(t)\rangle_J=
		\int^{z^*_f}_{z_i}\!\!\mathcal{D}\mu(z,z^*)
		\int^{z'^*_i}_{z'_f}\!\!\mathcal{D}\mu(z',z'^*)
		\int^{\phi^*_{kf}}_{\phi_{ki}}\!\!\mathcal{D}\mu(\phi_k,\phi^*_k)
		\int^{\phi'^*_{ki}}_{\phi'_{kf}}\!\!\mathcal{D}\mu(\phi'_k,\phi'^*_k)\,
		e^{\Gamma_\mathrm{S}[z,z^*]	+\Gamma_\mathrm{B}[\phi_k,\phi^*_k]
		+\Gamma^*_\mathrm{S}[z',z'^*]+\Gamma^*_\mathrm{B}[\phi'_k,\phi'^*_k]}
		\\
		O(z'_f,z^*_f)\rho_0(z_i,z'^*_i,\phi_{ki},\phi'^*_{ki})
		e^{i(\mathcal{S}[z,z^*,\phi_k,\phi^*_k]-\mathcal{S}^*[z',z'^*,\phi'_k,\phi'^*_k])}.
	\end{multline}
\end{widetext}

Here the limits of the integrals were written out explicitly for the sake of clarity, but they are understood to be integrated over. The notation $O(z'_f,z^*_f)$ and $\rho_0(z_i,z'^*_i,\phi_{ki},\phi'^*_{ki})$ indicates that the operators are expanded in the coherent-state basis $\ket{z,\phi_k}$ (we did not write any dependence on the bosonic variables for $\hat{O}$ because it is assumed to be of the form $\hat{O}_\mathrm{S}\otimes\mathbb{I}_\mathrm{B}$). The two sets of fields are the dynamical variables of the two independent path integrals representing the time evolution of the `ket' part ($z$, $z^*$, $\phi_k$, $\phi^*_k$) and of the `bra' part ($z'$, $z'^*$, $\phi'_k$, $\phi'^*_k$) of the density matrix: the Hermitian-conjugate time evolution of the `bra' part also brings in a minus sign in front of the relevant action functional in the exponent. The two actions are
\begin{widetext}
	\begin{multline}\label{eq:ActionFull}
		\mathcal{S}[z,z^*,\phi_k,\phi^*_k]\coloneqq\int_0^t\!\!\mathrm{d}t'\,
		\left(\frac{i}{2}\frac{z^*(t')\dot{z}(t')-\dot{z}^*(t')z(t')}{1+|z(t')|^2}
		+\frac{\epsilon}{4}n_z(z(t'))+\frac{\Delta}{4}n_x(z(t'))\right.
		\\
		\left.+\int^\infty_0\!\!\mathrm{d}k\,\left(\frac{i}{2}\left(\phi^*_k(t')
		\dot{\phi}_k(t')-\dot{\phi}^*_k(t')\phi_k(t')\right)
		-\omega_k\phi^*_k(t')\phi_k(t')\right)
		+\frac{\lambda}{4}n_z(z(t'))\int^\infty_0\!\!\mathrm{d}k\,h(\omega_k)
		\left(\phi_k(t')+\phi^*_k(t')\right)\right)
	\end{multline}
\end{widetext}
and its complex conjugate $\mathcal{S}^*[z',z'^*,\phi'_k,\phi'^*_k]$, where in the effective free spin Hamiltonian $H\left(z(t'),z^*(t')\right)=-\frac{\epsilon}{4}n_z(z(t'))-\frac{\Delta}{4}n_x(z(t'))$ (see definition~\eqref{eq:H} in the main text) we have ignored the standard renormalization counterterm necessary for consistently integrating out the bath from~\eqref{eq:ExpValFullPI}~\cite{FeynmanVernonFunctional,CaldeiraLeggett}, which in our case is just a constant---it is proportional to $\sigma^2_z=\mathbb{I}_\mathrm{S}$---and cancels out in the exponent of the path integral. On the other hand, note that the phases accumulated in the two parts of the evolution add up. The replacements $\sigma_i\longrightarrow-\frac{1}{2}n_i(z(t'))$ in the spin Hamiltonian are the result of the prescription to compute $H\left(z(t'),z^*(t')\right)$ for spin $1/2$.

The path integral~\eqref{eq:ExpValFullPI} is Gaussian in the continuous bosonic variables, which may therefore be integrated out analytically, leaving a path integral in the spin fields alone with the primed and unprimed variables coupled inside the influence functional: this is analogous to the mixing caused by the partial trace over the bath degrees of freedom in the canonical formalism.

Following the standard rules of path integration, we carry out the Gaussian integral by completing the square and find the result
\begin{widetext}
	\begin{equation}\label{eq:ExpValPI}
		\langle \hat{O}(t)\rangle_J=\mathcal{N}\int^{z^*_f}_{z_i}\!\!\mathcal{D}\mu(z,z^*)
		\int^{z'^*_i}_{z'_f}\!\!\mathcal{D}\mu(z',z'^*)\,
		e^{\Gamma_\mathrm{S}[z,z^*]+\Gamma^*_\mathrm{S}[z',z'^*]}
		O(z'_f,z^*_f)\rho_0(z_i,z'^*_i)	e^{i(\mathcal{S}_\mathrm{S}[z,z^*]
		-\mathcal{S}^*_\mathrm{S}[z',z'^*])-\Phi[z,z^*,z',z'^*,J]};
	\end{equation}
\end{widetext}
the overall normalization constant $\mathcal{N}$, corresponding to the value of the Gaussian integral with the completed square, is irrelevant to our purposes and will be dropped from now on.

The bosonic boundary terms appearing in~\eqref{eq:ExpValFullPI} vanish if the initial state of the bath is thermal, because the bosonic thermal state $\rho_0(\phi_k,\phi^*_k,\beta)$ may be regarded as a path integral in imaginary time, with the endpoints fixed by the values of $\phi_k$ and $\phi^*_k$: these are the very values from which the left and right propagators start and the trace operation makes the other ends of the propagators meet, closing the so-called Keldysh contour over itself and effectively removing the boundaries at which such terms would emerge.

Moreover, in the new path integral~\eqref{eq:ExpValPI} only the initial density matrix of the spin appears: the initial state of the bath must be assumed Gaussian in order for this simple integration step to be possible, and all necessary information about it is stored in the functional $\Phi[z,z^*,z',z'^*,J]$.

Finally, in the path integral we now have the free spin actions $\mathcal{S}_\mathrm{S}[z,z^*]$ and $\mathcal{S}^*_\mathrm{S}[z,z^*]$ (defined as in~\eqref{eq:ActionFull} without the parts involving bosonic fields) in the exponent, and the Feynman--Vernon influence functional $e^{-\Phi[z,z^*,z',z'^*,J]}$, whose exponent for the case of a thermal initial state of the bosons reads
\begin{widetext}
	\begin{equation}
		\Phi[z,z^*,z',z'^*,J]\coloneqq\frac{\lambda^2}{16}
		\int_0^t\!\!\mathrm{d}t'\int_0^{t'}\!\!\mathrm{d}t''\,\big(n_z(z(t'))
		-n_z(z'(t'))\big)\big(\xi_J(t'-t'')n_z(z(t''))
		-\xi^*_J(t'-t'')n_z(z'(t''))\big),
	\end{equation}
\end{widetext}
where $\xi_J(t)$ is the bath correlation function defined in the main text.

\subsection{Derivation of the error bounds}

Since $\xi_J(t)$, and hence $\Phi[z,z^*,z',z'^*,J]$, is linear in $J$, the influence functional for a spectral density $J(\omega)=J_0(\omega)+\Delta J(\omega)$ can be written as a product:
\[
	\begin{split}
		e^{-\Phi[z,z^*,z',z'^*,J]}
		&=e^{-\Phi[z,z^*,z',z'^*,J_0]-\Phi[z,z^*,z',z'^*,\Delta J]}
		\\
		&=e^{-\Phi[z,z^*,z',z'^*,J_0]}e^{-\Phi[z,z^*,z',z'^*,\Delta J]}.
	\end{split}
\]

Therefore, we can think of $\langle\hat{O}(t)\rangle_J$ as an expectation value over a bath with spectral density $J_0(\omega)$ reweighted with the functional $e^{-\Phi[z,z^*,z',z'^*,\Delta J]}$.

Writing out only $e^{-\Phi[z,z^*,z',z'^*,\Delta J]}$ as a series and using Eq.~\eqref{eq:InfluenceSuperOperator} from the main text to return to the operator formalism, one immediately obtains the form~\eqref{eq:DeltaExpValSeries} for $\Delta\langle\hat{O}(t)\rangle$. A first obvious step towards bounding its magnitude is to apply the triangle inequality to it and bound each statistical average separately:
\begin{equation}\label{eq:Series}
	|\Delta\langle\hat{O}(t)\rangle|
	\leq\sum^\infty_{n=1}\frac{|\langle\hat{O}(t)\mathcal{T}
	(-\hat{\Phi}[\hat{h}_\mathrm{I},\hat{h}'_\mathrm{I},\Delta J])^n\rangle_{J_0}|}{n!}.
\end{equation}
Now we need to study the $n$th-order term in the sum. Defining
\[
	\begin{split}
		\Delta\xi(t)&\coloneqq\xi_J(t)-\xi_{J_0}(t)
		\\
		&=\xi_{\Delta J}(t)
	\end{split}
\]
and its complex conjugate $\Delta\xi^*(t)$, at first order we have
\begin{widetext}
	\begin{multline}
		\langle\hat{O}(t)\mathcal{T}(-\hat{\Phi}[\hat{h}_\mathrm{I},\hat{h}'_\mathrm{I},
		\Delta J])
		\rangle_{J_0}=-\int^t_0\!\!\mathrm{d}t'\int^{t'}_0\!\!\mathrm{d}t''\,
		\left(\Delta\xi(t'-t'')	\left(
		\mathrm{Tr}(\hat{O}(t)\hat{h}_\mathrm{I}(t')\hat{h}_\mathrm{I}(t'')\hat{\rho}_0)
		-\mathrm{Tr}(\hat{O}(t)\hat{h}_\mathrm{I}(t'')\hat{\rho}_0\hat{h}_\mathrm{I}(t'))
		\right)\right.
		\\
		\left.+\Delta\xi^*(t'-t'')\left(
		\mathrm{Tr}(\hat{O}(t)\hat{\rho}_0\hat{h}_\mathrm{I}(t'')\hat{h}_\mathrm{I}(t'))
		-\mathrm{Tr}(\hat{O}(t)\hat{h}_\mathrm{I}(t')\hat{\rho}_0\hat{h}_\mathrm{I}(t''))
		\right)\right).
	\end{multline}
\end{widetext}
Increasing $n$ (and ignoring the denominator $n!$ for the moment), each new power of $-\hat{\Phi}[\hat{h}_\mathrm{I},\hat{h}'_\mathrm{I},\Delta J]$ brings in its own independent time integrations, $\Delta\xi(t'-t'')$, $\Delta\xi^*(t'-t'')$ and two $\hat{h}_\mathrm{I}$ or $\hat{h}'_\mathrm{I}$ operators inside the correlation functions, placed to the left or to the right of $\hat{\rho}_0$ in all possible ways, which quadruple the number of trace terms. Hence, the $n$th-order term in the series has the form of a $2n$-fold time integral of a linear combination of $4^n$ trace terms, each coming in with a prefactor
\[
	\xi_{i_1\dots i_p}(t_1,\dots,t_n)\coloneqq\prod^p_{m=1}\Delta\xi(t_{i_m})
	\!\!\prod_{j\neq i_1,\dots, i_p}\!\!\Delta\xi^*(t_j)
\]
with $p\leq n$ and a sign depending on $p$ and the operator placement inside the trace. Therefore, to bound $|\langle\hat{O}(t)\mathcal{T}(-\hat{\Phi}[\hat{h}_\mathrm{I},\hat{h}'_\mathrm{I},\Delta J])^n\rangle_{J_0}|$, we take the absolute value inside the time integrals and apply the triangle inequality to the integrand to bound each trace term separately. Note that at  this stage the distinction between $\Delta\xi(t)$ and $\Delta\xi^*(t)$ no longer matters, because for all $p,r\leq n$ we have
\[
	\begin{split}
		&|\xi_{i_1\dots i_p}(t_1,\dots,t_n)\mathrm{Tr}(\hat{O}(t)\mathcal{T}
		(\hat{h}_\mathrm{I}^{n+r})\hat{\rho}_0\mathcal{T}^*(\hat{h}_\mathrm{I}^{n-r}))|
		\\
		&=|\xi_{i_1\dots i_p}(t_1,\dots,t_n)||\mathrm{Tr}(\hat{O}(t)\mathcal{T}
		(\hat{h}_\mathrm{I}^{n+r})\hat{\rho}_0\mathcal{T}^*(\hat{h}_\mathrm{I}^{n-r}))|
		\\
		&=|\xi_{1\dots n}(t_1,\dots,t_n)||\mathrm{Tr}(\hat{O}(t)\mathcal{T}
		(\hat{h}_\mathrm{I}^{n+r})\hat{\rho}_0\mathcal{T}^*(\hat{h}_\mathrm{I}^{n-r}))|
		\\
		&=\prod^n_{i=1}|\Delta\xi(t_i)||\mathrm{Tr}(\hat{O}(t)\mathcal{T}
		(\hat{h}_\mathrm{I}^{n+r})\hat{\rho}_0\mathcal{T}^*(\hat{h}_\mathrm{I}^{n-r}))|,
	\end{split}
\]
where the notation $\hat{h}_\mathrm{I}^{m}$ is a shorthand for $m$ $\hat{h}_\mathrm{I}(t)$ operators at different times and $\mathcal{T}^*$ denotes inverse time ordering, in accordance with the Hermitian-conjugate time evolution of the right part of $\hat{\rho}_0$.

The operator traces $|\mathrm{Tr}(\hat{O}(t)\mathcal{T}(\hat{h}_\mathrm{I}^{n+r})\hat{\rho}_0\mathcal{T}^*(\hat{h}_\mathrm{I}^{n-r}))|$ can be bounded using the singular-value decomposition: defining the operator norm
\[
	||\hat{O}||\coloneqq||\hat{O}||_\infty=\sigma_1(\hat{O}),
\]
where $\sigma_1(\hat{O})$ is the highest singular value of the operator $\hat{O}$, we have
\[
	|\mathrm{Tr}(\hat{O}(t)\mathcal{T}(\hat{h}_\mathrm{I}^{n+r})\hat{\rho}_0\mathcal{T}^*
	(\hat{h}_\mathrm{I}^{n-r}))|\leq||\hat{O}||\,||\hat{h}_\mathrm{I}||^{2n}
\]
since $||\hat{\rho}_0||\leq1$. This bound is independent of both the positions and the time arguments of the interaction operators $\hat{h}_\mathrm{I}(t_i)$ inside the trace, so it can be factored out of the whole expression, leaving a sum of $4^n$ identical integrals:
\begin{equation}
	\begin{split}
		&|\langle\hat{O}(t)\mathcal{T}
		(-\hat{\Phi}[\hat{h}_\mathrm{I},\hat{h}'_\mathrm{I},\Delta J])^n\rangle_{J_0}|
		\\
		&\leq||\hat{O}||\,||\hat{h}_\mathrm{I}||^{2n}4^n\left(
		\int^t_0\!\!\mathrm{d}t'\int^{t'}_0\!\!\mathrm{d}t''\,|\Delta\xi(t'-t'')|\right)^n
		\\
		&=||\hat{O}||\left(\lambda^2
		\int^t_0\!\!\mathrm{d}t'\int^{t'}_0\!\!\mathrm{d}t''\,|\Delta\xi(t'-t'')|\right)^n,
	\end{split}
\end{equation}
where we have used $||\hat{h}_\mathrm{I}||^2=\frac{\lambda^2}{4}$. Summing all the terms of the series~\eqref{eq:Series}, one therefore arrives at the bound stated in Eq.~\eqref{eq:GeneralBound} of the main text.

\section{\label{sec:Examples}Results for some common spectral densities}

We have computed the correlation functions corresponding to several common spectral densities (which may just as well be used as variations), and analyzed their compliance with condition~\eqref{eq:StrongBoundCondition} in the main text, which ensures that the strongest of our two bounds Eq.~\eqref{eq:StrongBound} applies. Here we show our findings.

\subsubsection*{Ohmic spectral density}

For an ohmic spectral density with an exponential cutoff $J(\omega)=\pi\omega e^{-\omega/\Omega}$, the correlation function is
\begin{equation}
	\xi(t)=\frac{\psi^{(1)}\left(\frac{1+i\Omega t}{\beta\Omega}\right)
	+\psi^{(1)}\left(\frac{1-i\Omega t}{\beta\Omega}\right)}{\beta^2}
	+\frac{\Omega^2}{(\Omega t-i)^2},
\end{equation}
where $\psi^{(n)}(z)\coloneqq\frac{\mathrm{d}^n}{\mathrm{d}z^n}\frac{\Gamma'(z)}{\Gamma(z)}$ is the polygamma function of order $n$. This correlation function is absolutely integrable, because it has no singularities at any positive $t$ and its real and imaginary parts fall off at infinity as $\frac{1}{t^2}$ and $\frac{1}{t^3}$, respectively. Therefore, modifying a given spectral density by adding or removing an ohmic contribution---or changing the coefficient of an existing one---affects expectation values by introducing an error bounded by~\eqref{eq:StrongBound}.

\subsubsection*{Superohmic spectral densities}

The correlation function for superohmic densities with integer exponents $J(\omega)=\pi\omega^n e^{-\omega/\Omega}$ can easily be determined by differentiating the ohmic result with respect to $-\frac{1}{\Omega}$: for any given $n$ we have
\begin{equation}
	\xi(t)=\frac{\psi^{(n)}\left(\frac{1+i\Omega t}{\beta\Omega}\right)
	+\psi^{(n)}\left(\frac{1-i\Omega t}{\beta\Omega}\right)}{(-\beta)^{n+1}}	
	-n!\left(\frac{-i\Omega}{\Omega t-i}\right)^{n+1},
\end{equation}
where the temperature-dependent term falls off as $\frac{1}{t^n}$ and the real and imaginary parts of the remaining term both decrease faster than this (proportionally to $\frac{1}{t^{n+1}}$ and $\frac{1}{t^{n+2}}$, with the $(-i)^{n+1}$ at the numerator switching the real and imaginary parts from one value of $n$ to the next). In general, since a higher $n$ makes the correlation function decay faster, spectral densities of this type also satisfy the condition for the stronger bound.

\subsubsection*{Subohmic spectral density}

We could not compute the full correlation function for a subohmic spectral density of the form $J(\omega)=\pi\sqrt{\omega}e^{-\omega/\Omega}$, but we derived the limiting expressions at zero and infinite temperature:
\begin{equation}
	\xi_{\beta\longrightarrow\infty}(t)=\sqrt{\frac{\Omega^3}{\pi}}
	\frac{e^{i\frac{3}{4}\arctan(\Omega t)}}{2(1+\Omega^2t^2)^\frac{3}{4}},
\end{equation}
\begin{multline}
	\xi_{\beta\longrightarrow 0}(t)=\frac{1}{\beta}\sqrt{\frac{2\Omega}{\pi}
	\frac{1+\sqrt{1+\Omega^2t^2}}{1+\Omega^2t^2}}
	\\
	+i\sqrt{\frac{\Omega^3}{\pi}}
	\frac{\sin\left(\frac{3}{4}\arctan(\Omega t)\right)}{2(1+\Omega^2t^2)^\frac{3}{4}}.
\end{multline}

At zero temperature, the time decay is sufficiently fast for condition~\eqref{eq:StrongBoundCondition} to hold, while the infinite-temperature case gives a slow fall-off proportional to $\frac{1}{\sqrt{t}}$; however, since the correlation function does decay, the double time integral $\int^t_0\!\!\mathrm{d}t'\int^{t'}_0\!\!\mathrm{d}t''\,|\xi(t'-t'')|$ must scale slower than $t^2$, so it would be possible in principle to derive an intermediate bound for this case, at least numerically.

\subsubsection*{Single Dirac-delta mode}

Changing a given spectral density by adding or subtracting a mode with a specific frequency is a common practice in phenomenological modeling. However, when the mode is undamped, i.~e.\ a Dirac delta centered at some frequency $\omega_0$, this obviously yields a correlation function which oscillates indefinitely. In this case, only the bound~\eqref{eq:WeakBound} applies.

\subsubsection*{Single antisymmetrized Lorentzian mode}

It is often more realistic to treat individual modes using antisymmetrized Lorentzian peaks of the form $J(\omega)=\frac{\omega}{((\omega+\Omega)^2+\Gamma^2)((\omega-\Omega)^2+\Gamma^2)}$, corresponding to damped harmonic oscillators coupled to the central system. This both gives them a finite width (and hence dissipation properties) and provides an extremely versatile tool for constructing structured spectral densities to fit experimental data or simulate complex baths, as discussed in the main text and references therein.

For such a spectral density, the correlation function takes the form
\begin{equation}\label{eq:LorentzianCorrelation2}
	\begin{split}
		\xi(t)=&-\frac{2}{\beta}\sum^\infty_{k=1}
		\frac{\nu_ke^{-\nu_kt}}{(\Omega^2+\Gamma^2-\nu^2_k)^2+4\Omega^2\nu^2_k}
		\\
		&+\frac{e^{-\Gamma t}}{8\Omega\Gamma}\left(\coth\left(
		\frac{\beta}{2}(\Omega+i\Gamma)\right)e^{i\Omega t}\right.
		\\
		&\left.+\coth\left(\frac{\beta}{2}(\Omega-i\Gamma)\right)e^{-i\Omega t}
		+2i\sin(\Omega t)\right),
	\end{split}
\end{equation}
where $\nu_k\coloneqq\frac{2\pi k}{\beta}$ are the Matsubara frequencies, and decays exponentially, satisfying~\eqref{eq:StrongBoundCondition} because the series in~\eqref{eq:LorentzianCorrelation2} converges after integration.

\bibliography{References}
\bibliographystyle{unsrt}

\end{document}